\documentclass[10pt]{article}
\usepackage{graphicx}
\usepackage{lineno}
\usepackage{amsmath}
\usepackage{mathrsfs}

\usepackage{ifthen} 
\newboolean{uprightparticles}
\setboolean{uprightparticles}{false}
\usepackage{xspace} 
\usepackage{upgreek}

\def\CP                {{\ensuremath{C\!P}}\xspace}
\def\Title#1{\begin{center} {\Large #1 } \end{center}}
\def\Author#1{\begin{center}{ \sc #1} \end{center}}
\def\Address#1{\begin{center}{ \it #1} \end{center}}

\newcommand\pubblock{\rightline{\begin{tabular}{l} Proceedings of the Fifth Annual LHCP\\ \pubnumber\\
         \pubdate  \end{tabular}}}

\newenvironment{Abstract}{\begin{quotation} \begin{center} 
             \large ABSTRACT \end{center}\bigskip 
      \begin{center}\begin{large}}{\end{large}\end{center} \end{quotation}}

\newenvironment{Presented}{\begin{quotation} \begin{center} 
             PRESENTED AT\end{center}\bigskip 
      \begin{center}\begin{large}}{\end{large}\end{center} \end{quotation}}

\def\beq{\begin{equation}}
\def\eeq#1{\label{#1}\end{equation}}
\def\eeqn{\end{equation}}

\def\beqa{\begin{eqnarray}}
\def\eeqa#1{\label{#1}\end{eqnarray}}
\def\eeqan{\end{eqnarray}}

\let\bar=\overbar

\def\Dslash{\not{\hbox{\kern-4pt $D$}}}
\def\dslash{\not{\hbox{\kern-2pt $\del$}}}

\def\msb{{\bar{\ssstyle M \kern -1pt S}}}

\usepackage{color}

\newcommand\pubnumber{ LHCb-PROC-2017-031 }

\newcommand\pubdate{\today}

\def\affiliation{
On behalf of the LHCb Collaboration, \\
Center of High Energy Physics \\
Tsinghua University, Beijing, 100084, China}

\begin{document}

\large
\begin{titlepage}
\pubblock

\vfill
\Title{  Measurements of the  \CP violating phase $\phi_s$ at LHCb }
\vfill

\Author{ MENGZHEN WANG  }
\Address{\affiliation}
\vfill
\begin{Abstract}

The measurement of the mixing-induced \CP-violating phase $\phi_s$ in the  $B^0_s - \bar{B}^0_s$  system is one of the key goals of the LHCb experiment, and provides an excellent opportunity to search for New Physics beyond Standard Model. Using Run-I data, $\phi_s$ has been measured in several decay channels. These proceedings briefly summarise the previous LHCb Run-I results of $\phi_s$ measurement, and show the most recent results obtained by analysing the decay channel $B^0_s \to J/\psi K^+ K^-$  in the $K^+ K^-$ mass region above the $\phi$(1020) resonance.

\end{Abstract}
\vfill

\begin{Presented}
The Fifth Annual Conference\\
 on Large Hadron Collider Physics \\
Shanghai Jiao Tong University, Shanghai, China\\ 
May 15-20, 2017
\end{Presented}
\vfill
\end{titlepage}
\def\thefootnote{\fnsymbol{footnote}}
\setcounter{footnote}{0}

\normalsize

\section{Introduction}

In the Standard Model (SM) of particle physics, Charge-Parity (\CP) violation is one of the necessary conditions to explain the baryon asymmetry of the universe \cite{Sakharov:1967dj}. In the well-known Cabibbo-Kobayashi-Maskawa (CKM) formalism \cite{Kobayashi:1973fv}, the couplings between quarks of different flavors are summarised in the CKM matrix, which is a 3 $\times$ 3 unitary matrix:
\begin{equation}
\begin{bmatrix}
d' \\
s' \\
b'
\end{bmatrix}
= 
\begin{bmatrix}
     V_{ud} & V_{us} & V_{ub}\\
     V_{cd} & V_{cs} & V_{cb}\\ 
     V_{td} & V_{ts} & V_{tb}
\end{bmatrix}
\begin{bmatrix}
d\\
s\\
b
\end{bmatrix}
.
\end{equation}
In the SM, \CP violation arises through the complex phase in the CKM matrix. The \CP parameter $\phi_s$ is related to the interference of $B_s^0$ mixing and decay amplitudes \cite{Zhang:2012sq}.
The phase $\phi_s$ is very sensitive to New Physics, since its prediction in the SM is very precise: $\phi_s^{SM} = -2 arg(-\frac{V_{ts}V_{tb}^*}{V_{cs} V_{cb}^*}) = -36.5 ^{+1.3}_{-1.2}$ mrad \cite{Charles:2015gya}, where the small effect of the sub-leading corrections from the penguin amplitudes is ignored \cite{Fleischer:2015mla, Aaij:2014vda, Aaij:2015mea}. 

The LHCb experiment at the Large Hadron Collider (LHC) is designed to make precision measurements of \CP violation and rare decays of $b$ and $c$ hadrons \cite{Alves:2008zz}. The LHCb detector is a single-arm and forward spectrometer. With an excellent ability of tracking reconstruction and decay vertex location, the LHCb detector is an ideal detector for studying the long-lived heavy flavor particles.

\section{Measurement of $\phi_s$  at the LHCb experiment}
The measurement of $\phi_s$ is experimentally accessed via the following time-dependent asymmetry \cite{Zhang:2012sq}, where both $B^0_s$ and $\bar{B}^0_s$ decay to final state $f$:

\begin{equation}
\mathscr{A}_{CP} (t) = \frac{\Gamma_{\bar{B}^0_s}(t)-{\Gamma_{B^0_s}(t)}}{\Gamma_{\bar{B}^0_s}(t) + {\Gamma_{B^0_s}}(t)} = \frac{S_f \sin(\Delta m_s t) - C_f \cos(\Delta m_s t)}{\cosh(\frac{\Delta \Gamma_s  t }{2})  + A^{\Delta \Gamma} \sinh(\frac{\Delta \Gamma_s t}{2})},
\end{equation}
where $C_f$ parameterizes the direct \CP asymmetry,  $S_f$ and $A^{\Delta \Gamma}$ parameterize the mixing induced \CP asymmetries, related to $\phi_s$ by
\begin{equation}
S_f = \eta_f \sin\phi_s, A^{\Delta \Gamma} = -\eta_f \cos \phi_s.
\end{equation}
In the formulae above, $t$ is the decay time, $\eta_f$ is the \CP eigenvalue of the final state, $\Gamma_{B^0_s(\bar{B}^0_s)}$ is the time-dependent decay rate, $\Delta \Gamma_s \equiv \Gamma_L - \Gamma_H$ is the difference of the decay width between the two mass eigenstates, $B_L$ and $B_H$, and $\Delta m_s \equiv m_H - m_L$ is the mass difference. The parameters of \CP violation are obtained experimentally thanks to a flavour-tagged time-dependent angular analysis.

\subsection{Previous LHCb Run-I results}
Using the data collected in 2011 and 2012, with a center of mass energy of 7 and 8 TeV, respectively, and an integrated luminosity of 3fb$^{-1}$, the LHCb collaboration has measured $\phi_s$ in several different decay channels of $B_s^0$ mesons. The analysis of  $B_s^0 \to J/\psi K^+ K^-$ decays, with the invariant mass of $K^+ K^-$ in $\phi$(1020) region, shows that $\phi_s = -0.058 \pm 0.049 \pm 0.006$ rad. In this analysis, $\phi_s$ is also measured for the first time independently for each polarization. \cite{Aaij:2014zsa}.  A flavour-tagged time-dependent angular analysis of $B_s^0 \to J/\psi \pi^+ \pi^-$ decay shows that $\phi_s = +0.070 \pm 0.068 \pm 0.008$ rad \cite{Aaij:2014dka}. An analysis of the time evolution of the $\bar{B}_s^0 \to D_s^+ D_s^-$ decay gives a result of $\phi_s = +0.02 \pm 0.17 \pm 0.02$ rad \cite{Aaij:2014ywt}. The analysis of $B_s^0 \to \psi(2S) \phi$ decays gives the first measurement of $\phi_s$ in a decay containing the $\psi(2S)$ resonance, and the result is $\phi_s =+ 0.23 ^{+0.29}_{-0.28} \pm 0.02$ rad \cite{Aaij:2016ohx}. All results are compatible and dominated by the result obtained using $B^0_s \to J/\psi K^+ K^-$ decays.

\subsection{Measurement in $B_s^0 \to J/\psi K^+ K^-$ above $\phi(1020)$ region}
 A flavour-tagged time-dependent angular analysis is performed to determine $\phi_s$ in the $B_s^0 \to J/\psi K^+ K^-$ decay, with the invariant mass of $K^+ K^-$ above $\phi(1020)$ region \cite{Aaij:2017zgz}.

The decay rates into the $J/\psi K^+ K^-$ final states are functions of proper time, 
\begin{equation}
\Gamma(t) \propto e^{\Gamma_s t} \left(\frac{|A|^2 + |\bar{A}|^2}{2} \cosh \frac{\Delta \Gamma_s t}{2} + \frac{|A|^2 - |\bar{A}|^2}{2}\cos(\Delta m_s t) - Re(A^*\bar{A})\sinh\frac{\Delta\Gamma_s t}{2} - Im(A^*\bar{A})\sin(\Delta m_s t) \right),
\end{equation}
\begin{equation}
\bar{\Gamma}(t) \propto e^{\Gamma_s t} \left(\frac{|A|^2 + |\bar{A}|^2}{2} \cosh \frac{\Delta \Gamma_s t}{2} + \frac{|A|^2 - |\bar{A}|^2}{2}\cos(\Delta m_s t) - Re(A^*\bar{A})\sinh\frac{\Delta\Gamma_s t}{2} + Im(A^*\bar{A})\sin(\Delta m_s t) \right),
\end{equation}
where $\Gamma_s \equiv \frac{\Gamma_L + \Gamma_H}{2}$ is the average width of two mass eigenstates. 
$A$($\bar{A}$) is the  total decay amplitude for a $B_s^0$($\bar{B}_s^0$) meson at decay time equal to zero. $A$ and $\bar{A}$ are taken to be sum over all the amplitudes of $K^+ K^-$ resonance with different  transversities  as well as the non-resonant amplitude,  labelled as $A_i$ and $\bar{A_i}$: $A = \sum A_i$, while $\bar{A} = \frac{q}{p} \bar{A_i} = \eta_i |\lambda_i| e^{-i\phi_s^i} \bar{A_i}$, where $\eta_i$ is the \CP eigenstate value of the final state. Thus the terms with $A^*\bar{A}$ are sensitive to $\phi_s$.

\begin{figure}[htb]
\centering
\includegraphics[height=0.9in]{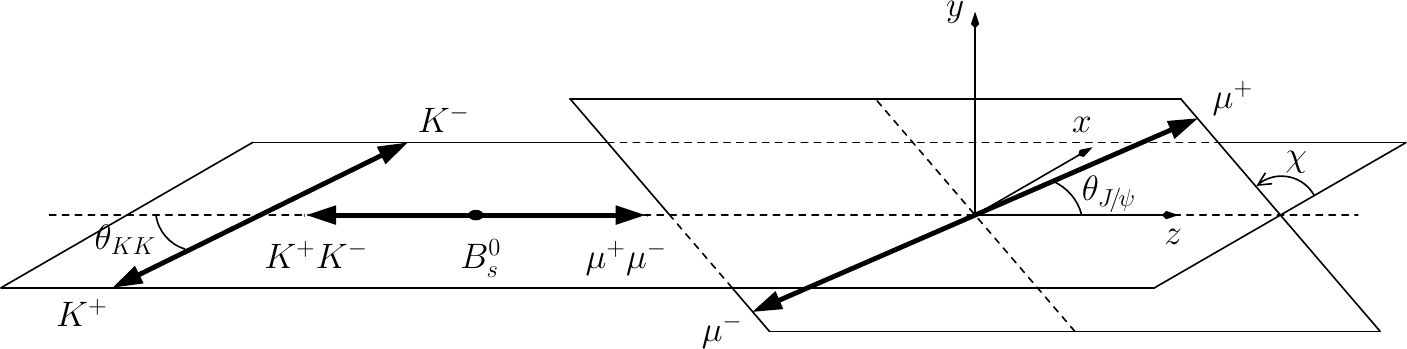}
\caption{Diagramatic representation of helicity angles of the decay $B_s^0 \to J/\psi K^+ K^-$.}
\label{fig:angle}
\end{figure}

The amplitudes are themselves functions of four variables: $m_{KK}$, the invariant mass of $K^+ K^-$, and three helicity-basis angular variables ($\cos \theta_{KK}$, $\cos \theta_{J/\psi}$ and $\chi$). These angles are shown in Fig.~\ref{fig:angle}.

As the definition of the total amplitude $A$ and $\bar{A}$ is related to the particle type of the $B$ meson at decay time equal to zero, and $B_s^0$ and $\bar{B}_s^0$ could oscillate to each other,  it's important to determine the flavour of the $B$ meson when it is generated. Two independent classes of flavour-tagging algorithms, the opposite-side(OS) tagger \cite{Aaij:2012mu}, as well as the same-side kaon(SSK) tagger \cite{Aaij:2016psi} are used for this purpose. The OS tagger method exploits the feature that the $b\bar{b}$ quark pairs are produced together in the proton-proton ($pp$) interaction, and the SSK  is a neural-networks-based algorithm, exploiting the $b$ hadron production mechanism in $pp$ collisions. The effective tagging power for the combined taggers, defined as $\epsilon_{tag}\langle (1-2\omega)^2 \rangle$ , is $(3.82 \pm 0.13\pm0.12)\%$, where $\epsilon_{tag}$ is the efficiency of the tagger, and $\omega$ corresponds to the mistag probability.

\begin{figure}[htb]
\centering
\includegraphics[height=1.1in]{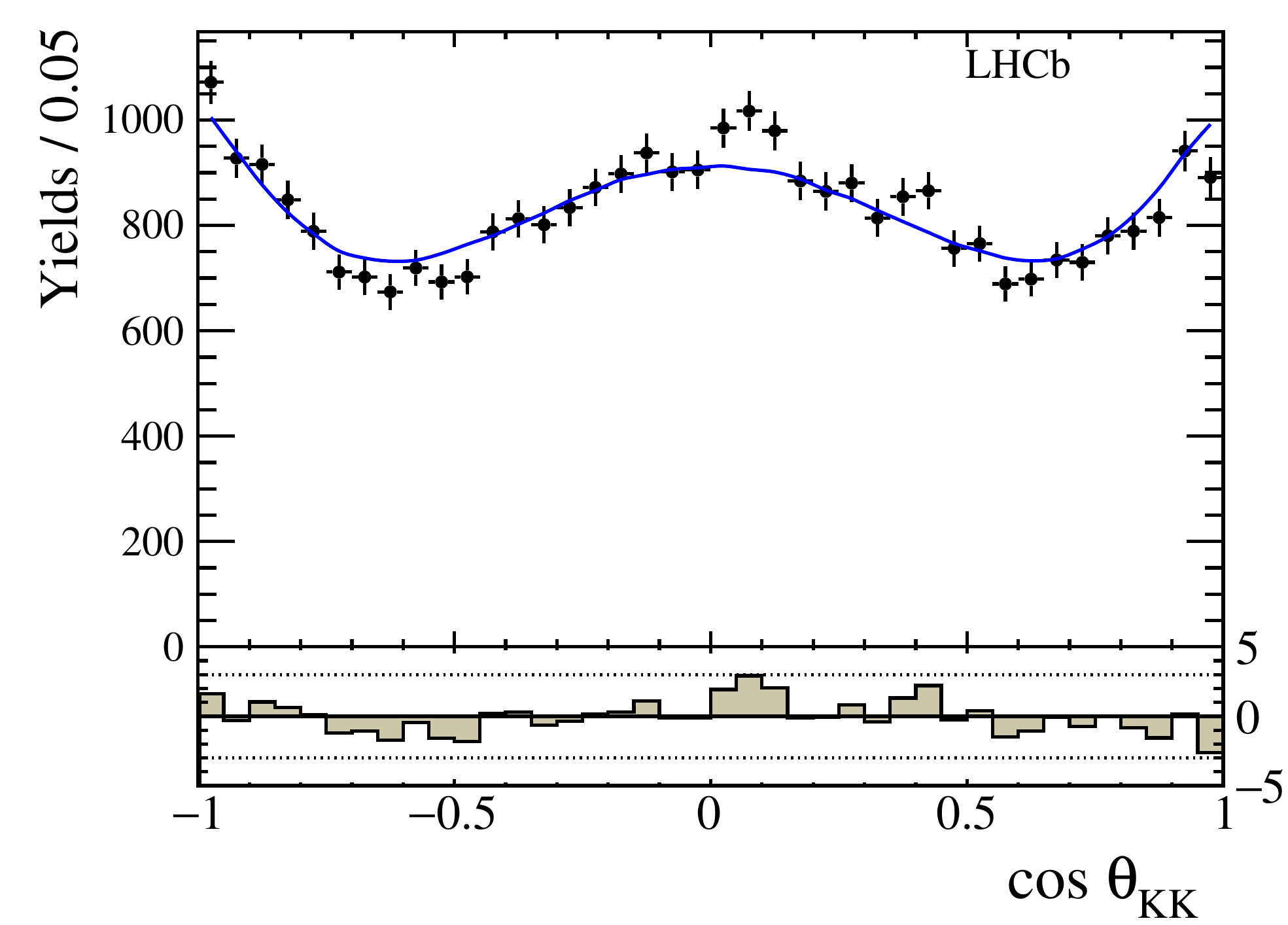}
\includegraphics[height=1.1in]{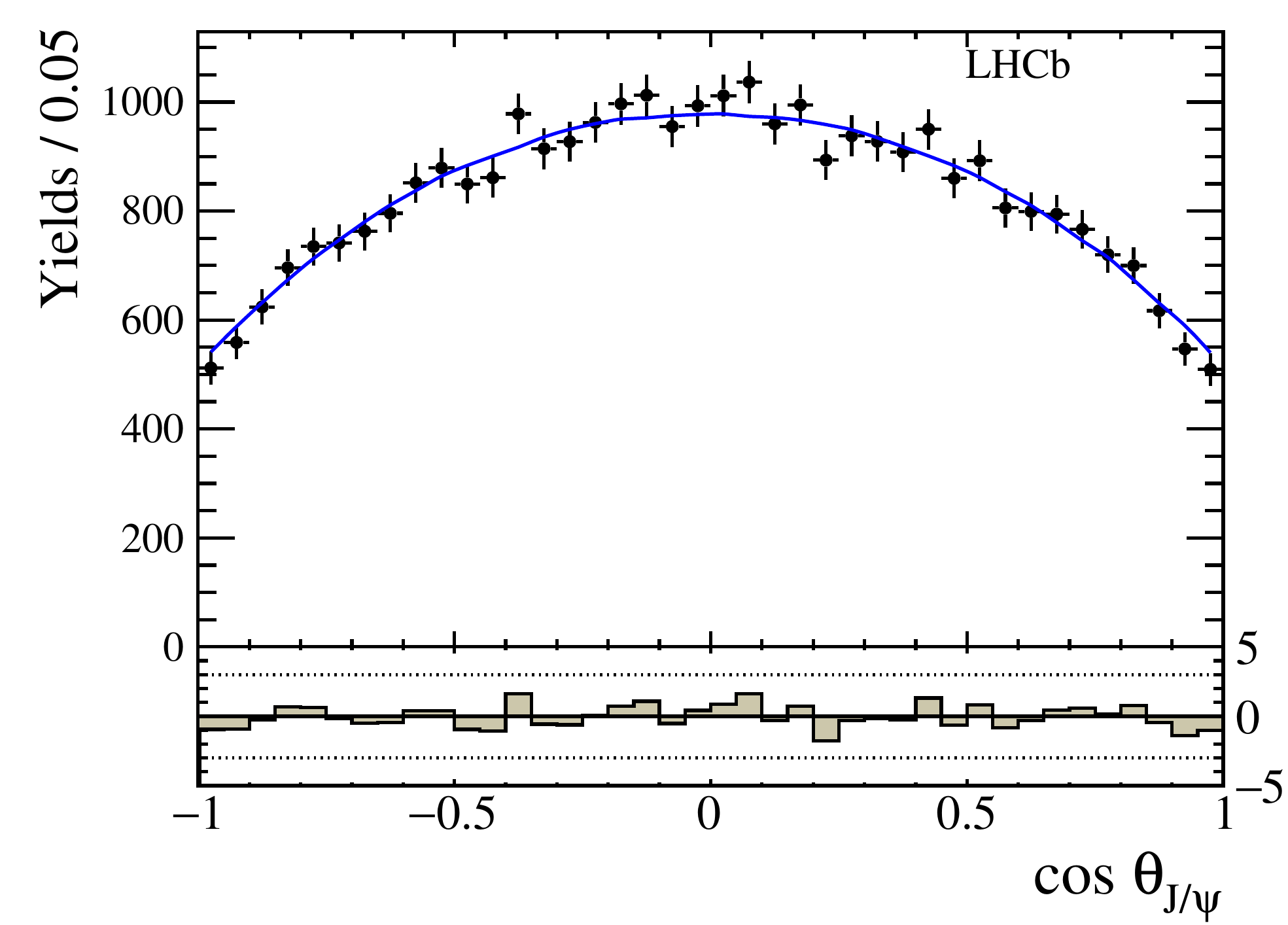}
\includegraphics[height=1.1in]{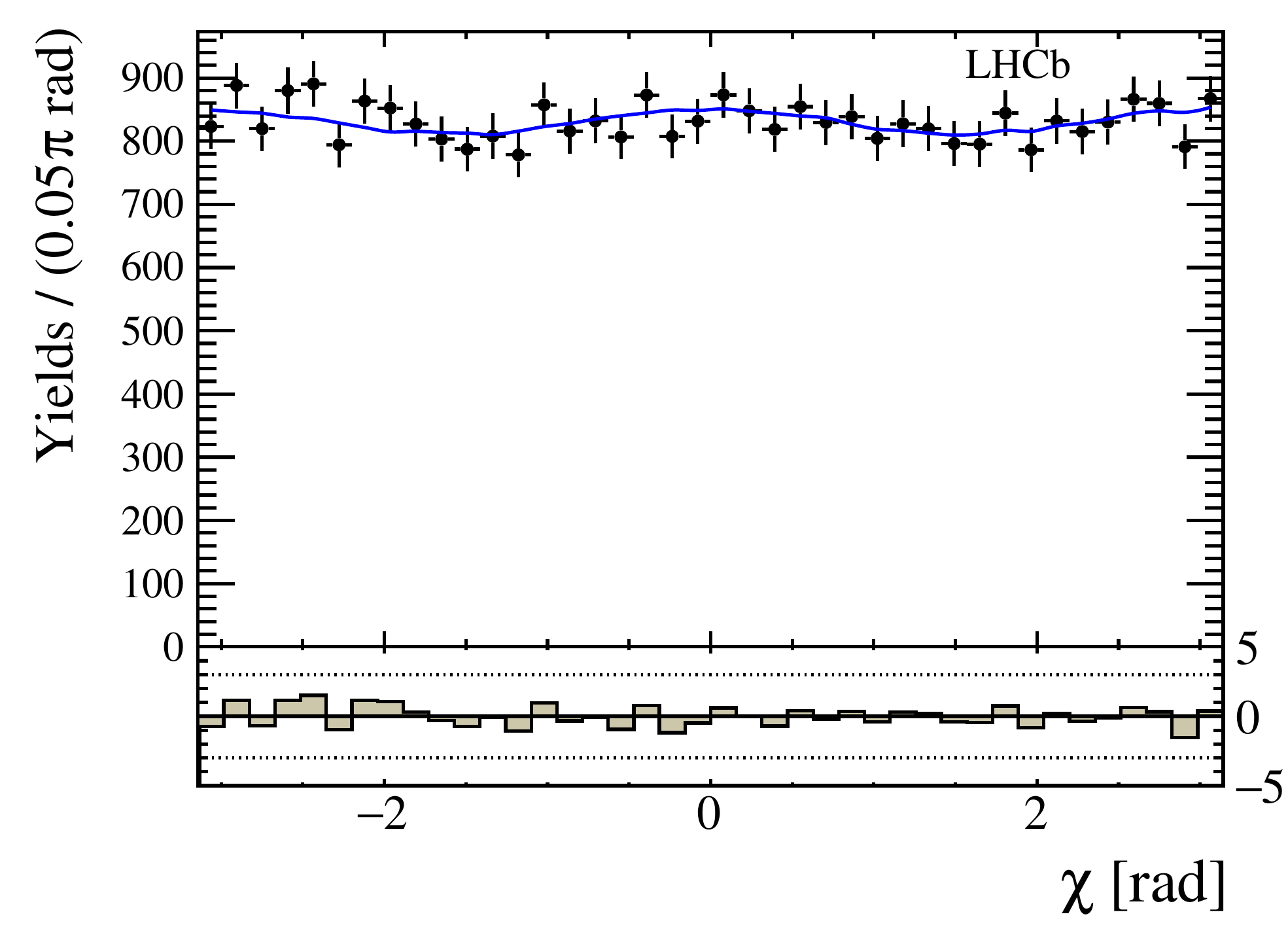}
\includegraphics[height=1.1in]{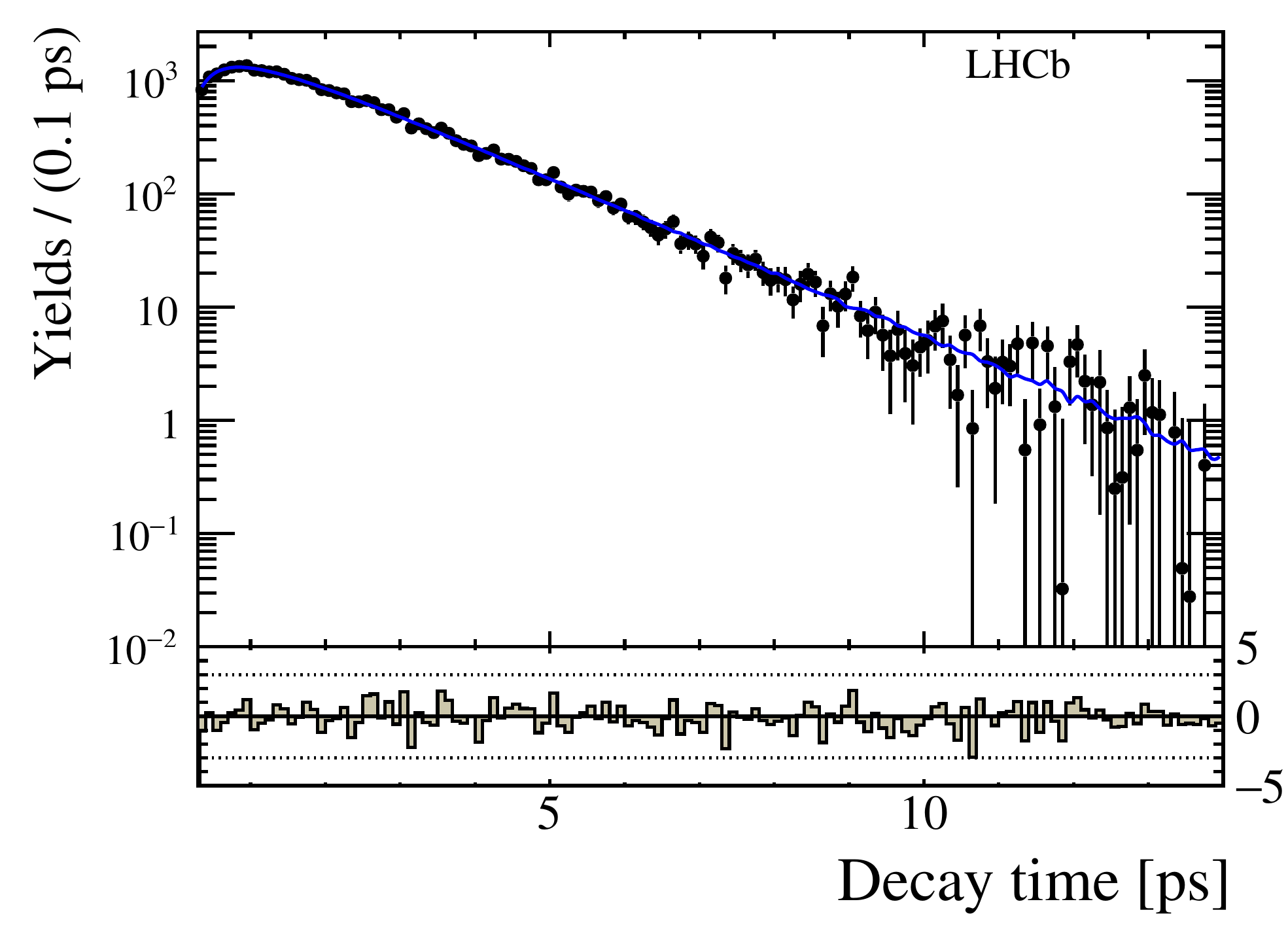}
\caption{Projection of the fit overlaid to data points in the $m_{KK}$ region above the $\phi(1020)$ resonance. The black points correspond to data, and the blue curve correspond to the fit results.  The difference between data and the fit result divided by the uncertainty is also shown at the bottom of each figure.}
\label{fig:projections1}
\end{figure}

\begin{figure}[htb]
\centering
\includegraphics[height=2in]{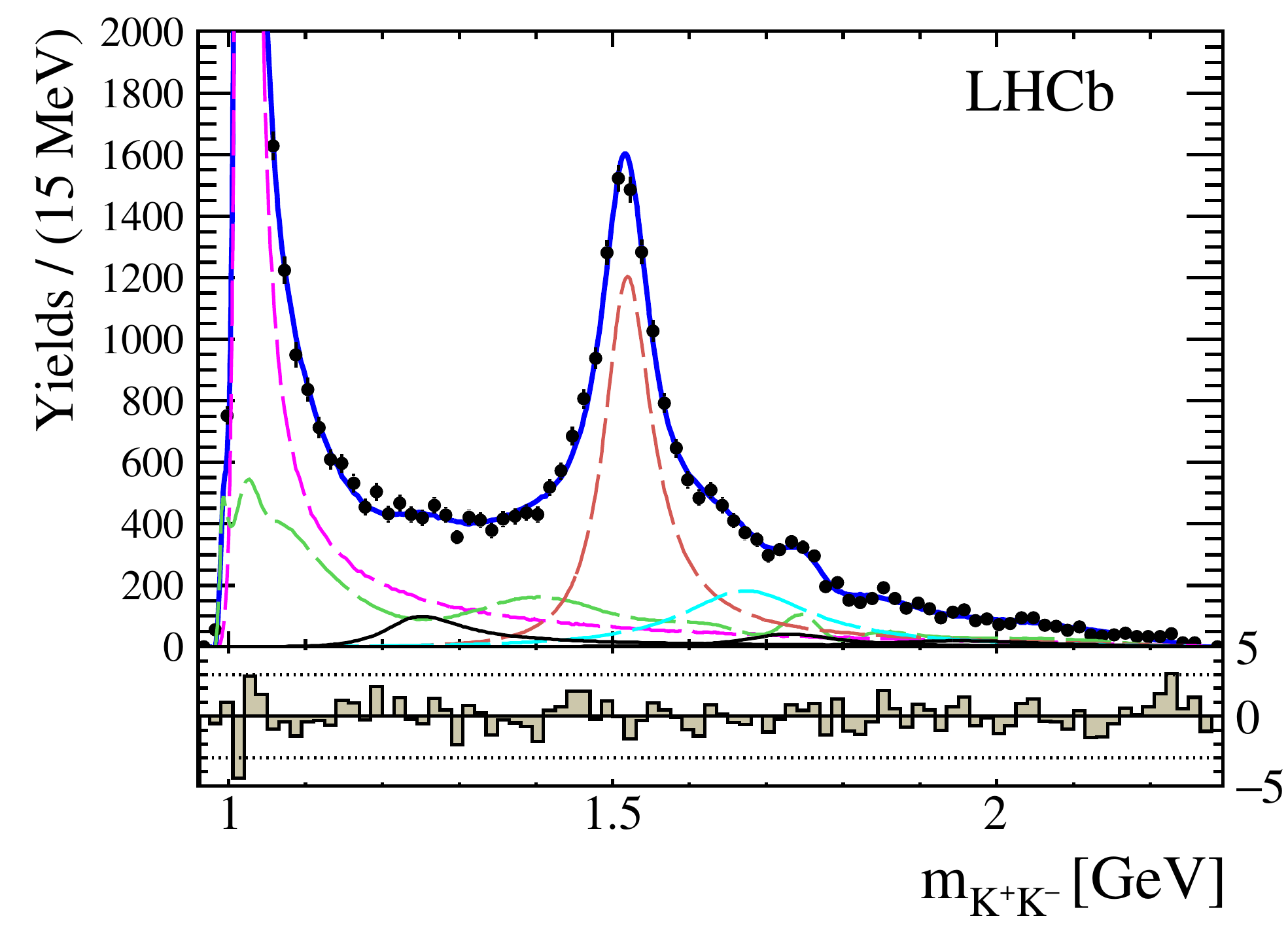}
\caption{Fit projection of $m_{KK}$. The points correspond to the data. The magenta curve, brown curve, cyan long-dashed curve, green long-dashed curve, black solid curve correspond to  $\phi$(1020),  $f_2'$(1525), $\phi$(1680), the S-wave component and other two $f_2$ resonances, respectively. The blue solid curve is the total fit result. The difference between data and the fit, divided by the uncertainty of each bin is shown at the bottom of the figure.}
\label{fig:projections2}
\end{figure}

The \CP parameters are obtained from a maximum likelihood fit  which is performed with a signal-only probability density function (PDF) of five-dimensional distributions: $B_s^0 (\bar{B}_s^0)$ decay time, $m_{KK}$ and helicity angles. 
The fit results can be seen in Fig.~\ref{fig:projections1}  and  Fig.~\ref{fig:projections2}. The fit results for these variables in high $m_{KK}$ region show that $\phi_s = 119 \pm 107 \pm 34$ mrad, $\Gamma_s = 0.650 \pm 0.006 \pm 0.004$ ps$^{-1}$, and $\Delta \Gamma_s = 0.066 \pm 0.018 \pm 0.010$ ps$^{-1}$.

In the mass spectrum of $K^+ K^-$, the resonance structure of $f_2'$(1525) is observed, with M = $1522.2 \pm 1.3 \pm 1.1$ MeV/$c^2$, and $\Gamma$ = $78.0 \pm 3.0 \pm 3.7$ MeV/$c^2$.

\section{Conclusions}
\begin{figure}[htb]
\centering
\includegraphics[height=2.0in]{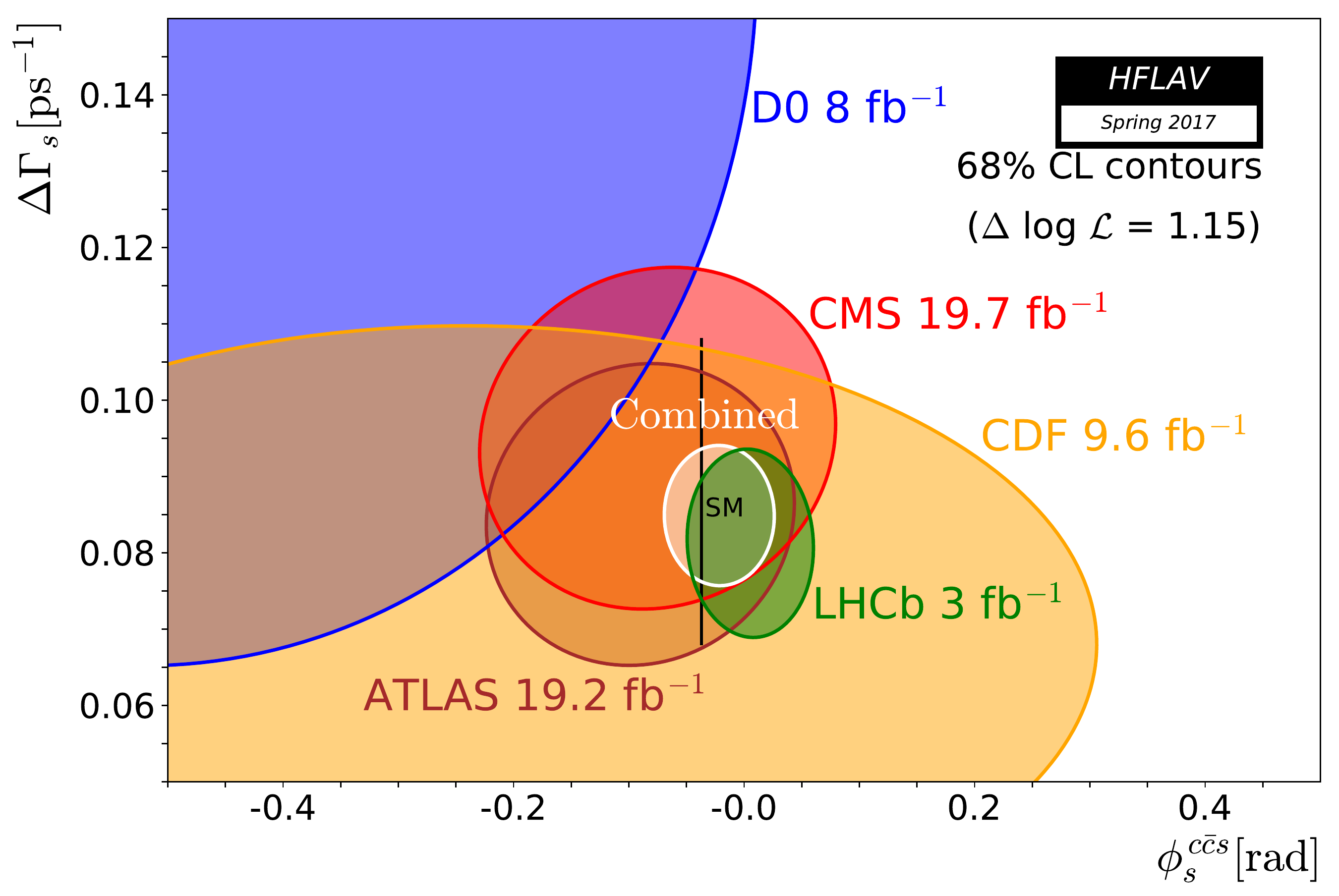}
\caption{The average results of $\phi_s$ measurement on each experiments. The result of the LHCb experiment is in agreement with the SM prediction.}
\label{fig:ave}
\end{figure}

With an excellent ability of tracking and vertex location, as well as the large $b$ and $c$ cross sections in LHC, the LHCb experiment plays an important role in searching for New Physics in \CP violation of heavy flavor particles, and the measurement of $\phi_s$ provides an approach to this goal. The previous LHCb Run-I results from the analyses of $B_s^0 \to J/\psi \phi$, $B_s^0 \to J/\psi \pi^+ \pi^-$, $B_s^0 \to D_s^+ D_s^-$ and $B_s^0 \to \psi(2S) \phi$ are consistent with the Standard Model prediction \cite{Aaij:2014zsa, Aaij:2014dka, Aaij:2014ywt, Aaij:2016ohx}. The latest measurement of $\phi_s$ is performed with the $B_s^0 \to J/\psi K^+ K^-$ decay, with the invariant mass of $K^+ K^-$ above the $\phi$ region. With a flavour-tagged time-dependent angular analysis, we determine 
\begin{center}
$\phi_s = 119 \pm 107 \pm 34$ mrad\\
$\Gamma_s =  0.650 \pm 0.006 \pm 0.004$ ps $^{-1}$\\
$\Delta \Gamma_s = 0.066 \pm 0.018 \pm 0.010$ ps $^{-1}$.
\end{center}
This new result of $\phi_s$ is also in agreement with the SM prediction $-36.5^{+1.3}_{-1.2}$ mrad \cite{Charles:2015gya}. The average result in the LHCb experiment can be found in Fig.~\ref{fig:ave} \cite{Amhis:2016xyh}.

\bibliographystyle{plain}
\bibliography{Ref}

\end{document}